\documentclass[english]{article}
\usepackage[T1]{fontenc}
\usepackage[latin9]{inputenc}
\usepackage{graphicx}

\makeatletter
\newcommand{\lyxaddress}[1]{
	\par {\raggedright #1
	\vspace{1.4em}
	\noindent\par}
}

\makeatother

\usepackage{babel}
\begin{document}
\title{Acrobot Swing Up with MATLAB}
\author{Yu Xiao}
\maketitle

\lyxaddress{\begin{center}
Boston University, Boston, MA
\par\end{center}}

\section{The Acrobot}

The acrobot usually refers to a 2-link underactuated robot with one
end fixed at some point in the coordinate frame. It looks similar
to a 2-link robot arm, but with only one actuator at the elbow. This
feature makes it an underactuated robot. Underactuated robot is a
family of robots which draws particular interests among engineers.
They are hard to control since they are usually highly non-linear.
The acrobot is one of the simplest examples in this family, as its
equations of motion is obtainable by manual derivation.

In this note, we focus on the classical swing up task for acrobot.
The task aims at bring the acrobot from the bottom position to the
up right position. We present solutions for this task using an optimization
based method with manually derived system equations. We studied two
kind of acrobot: the classical 2-link acrobot and an extended version
with three links. The computer programming environment we used in
the experiments is MATLAB, with a open-sourced MATLAB library called
OptimTraj.

\section{Theoretical Preliminaries}

\subsection{State Space Model of Dynamical Systems}

People have been using differential equations to describe nature since
the establishment of morden science. These mathematical objects equating
variables and their derivatives have numerous application in many
different fileds. Some of the most celerbrated examples including
heat equation describing heat diffusion in rods, equations describing
the growth of population and our favorite equations describing the
motion of rigid bodies.

The state space model of dynamical system is a set of first-order
ordinary differential equation(ODE for short) describing the behaviour
of a dynamical system. Using a set of generalized coordinates, we
can reduce high-order differential equations into first-order.
\begin{center}
$\dot{x}=f(x(t),u(t),t)$,
\par\end{center}

is a typical state space model. $x$ denotes the state space variables,
$\dot{x}$ denotes their first-order derivatives, $u$ is the control
input. Control input isn't a necessity for every system. A passive
system is a system with no control input. A special form of ODE we
will use in is note is ODEs with a mass matrix.
\begin{center}
$M\dot{x}=f(x(t),u(t),t)$
\par\end{center}

The mass matrix $M$ raise when the function $f$ no longer directly
describe the derivatives of state space variables, but a weighted
sum of the them. This form of ODEs are common results of motion of
rigid bodies, where we directly acquire the mass matrix. We are going
to use this form in the numerical solution of passive systems. However,
we need to recover the standard form by calculating the inverse of
mass matrix.
\begin{center}
$\dot{x}=M^{-1}f(x(t),u(t),t)$
\par\end{center}

The state space model of a dynamical is mostly non-linear, which means
that the function $f(x,u)$ is usually non-linear. The classical tools
analyzing the behaviour of a non-linear systems is Lyapunov theory.
In this note, we want to emphasize the trajectory optimization method
for control, instead of the theoretical properties of the objective
robot. So we will leave theoretical analysis of the system to study
in the future.

\subsection{The General Optimal trajectory Design Problem}

Trajectory optimization is a class of optimal control methods. We
are going to give up sloving optimal feedback controller, and instead
using optimization tool to find a good solution for a single initial
condition. The dynamics of the system is
\begin{center}
$\dot{x}=f(x(t),u(t),t)$,
\par\end{center}

and initial and final conditions can be defined within some prescribed
lower and upper bounds
\begin{center}
$\psi_{i,l}\leq\psi_{i}(x(t_{i}),u(t_{i}),t_{i})\leq\psi_{i,u}$,
\par\end{center}

\begin{center}
$\psi_{f,l}\leq\psi_{f}(x(t_{f}),u(t_{f}),t_{f})\leq\psi_{f,u}$.
\par\end{center}

In addition, we can impose simple bounds on the state variables
\begin{center}
$x_{l}\leq x(t)\leq x_{u}$
\par\end{center}

and on control variables
\begin{center}
$u_{l}\leq u(t)\leq u_{u}$.
\par\end{center}

The basic problem is to determine the control vector $u(t)$ to minimize
the performance index
\begin{center}
$J=\phi(x(t_{f}),t_{f})$,
\par\end{center}

where $\phi$ is the pre-defined cost function. To solve the optimal
trajectory design problem, we form it into a non-linear programming(NLP)
problem. The general NLP problem requires finding the $n$ vectors
to solve
\begin{center}
$min_{x}F(x)$,
\par\end{center}

subject to $m$ constraints
\begin{center}
$c_{L}\leq c(x)\leq c_{U}$,
\par\end{center}

and bounds
\begin{center}
$x_{L}\leq x\leq x_{U}$,
\par\end{center}

where $F(x)$ is a scalar function, and $c(x)$ are non-linear constraints.

\subsection{Direct Collocation}

To solve the general optimal trajectory design problem with NLP solver,
we need to form a scalar cost function $J$ over trajectories, and
convert dynamic constraint into inequalities. According to method
taken in this step, trajectory optimization methods are often divided
into two classes: indirect method and direct method. In the direct
setting, we will first discretize the state variable $x$ and control
input $u$, then formulate an optimization over the trajector. In
this note, we use the direct method.

The idea of direct transcriping is to discritize time into intervals(usuallt
equal intervals) as
\begin{center}
$t_{i},t_{2},t_{3},\ldots,t_{f}$.
\par\end{center}

The states and controls can be discretized over the trajectory by
defining $x_{k}=x(t_{k})$ and $u_{k}=u(t_{k})$. The discretized
states and controls are used as decision variables of NLP problem,
which is
\begin{center}
$y=[x_{1},u_{1},\ldots,x_{N},u_{N}]$.
\par\end{center}

For the dynamic constraints(the equations of motion), we can form
an equality constraint on every interval
\begin{center}
$x_{k+1}-x_{k}-hf_{k}=0$,
\par\end{center}

where $h$ is the step size of discretization(usually fixed), and
$f_{k}=f(x_{k},u_{k},t_{k})$. We can use the technique of constructing
a defect variable for each of these equality constraint, as
\begin{center}
$\zeta_{k}=x_{k+1}-x_{k}-hf_{k}$.
\par\end{center}

As a result of the transcription, the optimal control constraints
can be replaced by a set of inequalities of decision variables
\begin{center}
$c_{L}\leq c(y)\leq c_{U}$
\par\end{center}

where
\begin{center}
$c(y)=[\zeta_{1},\zeta_{2},\ldots,\zeta_{N-1},\psi_{1},\psi_{N},]$,
\par\end{center}

and
\begin{center}
$c_{L}=[0,0,\ldots,0,\psi_{i,l},\psi_{f,l}]$,
\par\end{center}

\begin{center}
$c_{U}=[0,0,\ldots,0,\psi_{i,u},\psi_{f,u}]$.
\par\end{center}

Collocation is an important methods used to transcribe differential
dynamic constraints into a set of algebraic constraints, in order
to form an NLP problem. The basic idea is to choose a polynomial up
to a certain degree with a number of points with boundary values matched
control and state value at knots, and to enforce the polynomials to
satisfy the equations of motion ar the collocation points(intermediate
points between knots).

\section{Mathematical Notation}

There will be plenty of equations in this note, so it's good to clarify
mathematical notation before jumping into the equations. Our acrobot
lives in a physical world with a gravitational acceleration germ $g$.
It has two massless link, and two point mass. A figure depicting our
acrobot look as the follow.
\begin{center}
\includegraphics[height=150pt]{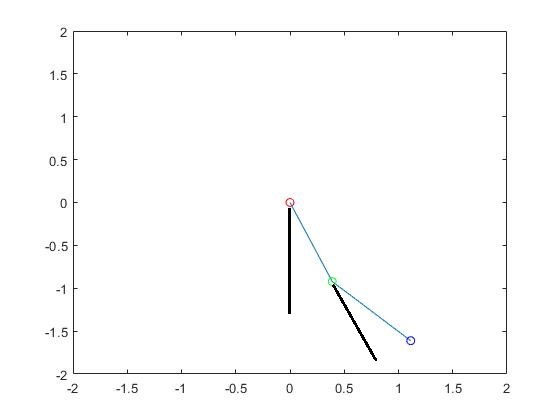}
\par\end{center}

In this figure, the two line segments in blue represent the massless
links. We will use $l_{1}$to to represent the length of the first
link(the upper one in this figure), and use $l_{2}$ for the length
of the second link(the lower one in this figure). The three small
circles in different colors(red, green and blue) are the ends of links.
The red point is the fixed end of the first link; the green point
is a revolut joint connecting two links with a mass $m_{1}$; the
blue point is a blue point is the end of the second link with a mass
$m_{2}$. The angle of the first link is represent by $q_{1}$, and
the second angle in represent by $q_{2}$. Since there are plenty
of trigonometry functions in the equations of motion, we use a short
hand for sinesoid functions. We use $c_{\theta}$ to denote function
$cos(\theta)$, and use $s_{\theta}$ to denote function $sin(\theta)$.
We use $\dot{q}$ to denote the first-order time derivative of $q$,
and use $\ddot{q}$ to denote the second time derivative of $q$.
$\tau$ denotes torques, from gravity or control input.

\section{Acrobot Swing Up}

\subsection{Equations of Motion}

The acrobot is a two-link pendulum, while only the second joint has
an actuator. Thus, it is an underactuated robot.

We will first derive the dynamics model of the double pendulum, and
simulate a passive one in MATLAB. We are going to use the method of
Lagrange to derive the equations of motion.
\begin{center}
$\frac{d}{dt}\frac{\partial L}{\partial\dot{q_{i}}}-\frac{\partial L}{\partial q_{i}}=Q_{i}$,
\par\end{center}

where L denotes the Lagrangian, $q_{i}$denotes the generalized coordinates
and Q$_{i}$ denotes the force applied.
\begin{center}
$L=T-U$
\par\end{center}

\begin{center}
$T=\frac{1}{2}(m_{1}+m_{2})l_{1}^{2}$($\dot{q_{1}}$)$^{2}$+$\frac{1}{2}m_{2}l_{2}^{2}(\dot{q_{1}}+\dot{q_{2}})^{2}+m_{2}l_{1}l_{2}\dot{q_{1}}(\dot{q_{1}}+\dot{q_{2}})c_{2}$
\par\end{center}

\begin{center}
$U=-(m_{1}+m_{2})gl_{1}c_{1}-m_{2}gl_{2}c_{1+2}$
\par\end{center}

Plugging the Lagrangian into the Lagrange Equations, we obtain the
equations of motion.

Then we can reform the equations of motion into a characteristic form
called the Manipulator Equations, which takes the form:
\begin{center}
$M(q)\ddot{q}+C(q,\dot{q})\dot{q}=\tau_{g}(q)+Bu$,
\par\end{center}

where q is the state vector, I is the inertia matrix, C captures Coriolis
forces, and $\tau_{g}$is the gravity vector. The matrix B maps control
inputs u into generalized force. In our first task to simulate the
passive double pendulum, we don't have to worry about B or u. The
other matrix take the following form:
\begin{center}
$M(q)=\left[\begin{array}{cc}
(m_{1}+m_{2})l_{1}^{2}+m_{2}l_{2}^{2}+2m_{2}l_{1}l_{2}c_{2} & m_{2}l_{2}^{2}+m_{2}l_{1}l_{2}c2\\
m_{2}l_{2}^{2}+m_{2}l_{1}l_{2}c_{2} & m_{2}l_{2}^{2}
\end{array}\right]$
\par\end{center}

\begin{center}
$C(q,\dot{q})=\left[\begin{array}{cc}
0 & -m_{2}l_{1}l_{2}(2\dot{q_{1}}+\dot{q_{2}})s_{2}\\
m_{2}l_{1}l_{2}\dot{q_{1}}s_{2} & 0
\end{array}\right]$
\par\end{center}

\begin{center}
$\tau_{g}(q)=-g\left[\begin{array}{c}
(m_{1}+m_{2})l_{1}s_{1}+m_{2}l_{2}s_{1+2}\\
m_{2}l_{2}s_{1+2}
\end{array}\right]$
\par\end{center}

\subsection{Dynamics Model of 2-link Acrobot}

Alright, we now have a horrific looking second-order differential
equation. The next necessary step is to reform the Manipulator Equations
into first-order differential equations, so that we can form a state
space model. The state space variable we are going to use is $q=\left[\begin{array}{cccc}
q_{1} & q_{2} & \dot{q_{1}} & \dot{q_{2}}\end{array}\right]$. The first-order differential equations we wish to have takes the
form:
\begin{center}
$Mass(q)\dot{q}=f(q)$
\par\end{center}

We use the matrix I and C to construct the Mass matrix:
\begin{center}
$Mass(q)=\left[\begin{array}{cc}
I_{2\times2} & 0_{2\times2}\\
C(q) & M(q)
\end{array}\right]$,
\par\end{center}

which is a 4-by-4 matrix. Then we can reformulate the Manipulator
Equation into a set of 4 differential equations:
\begin{center}
$\left[\begin{array}{cc}
I_{2\times2} & 0_{2\times2}\\
C(q) & M(q)
\end{array}\right]\left[\begin{array}{c}
\dot{q_{1}}\\
\dot{q_{2}}\\
\ddot{q_{1}}\\
\ddot{q_{2}}
\end{array}\right]=\left[\begin{array}{c}
\dot{q_{1}}\\
\dot{q_{2}}\\
-g(m_{1}+m_{2})l_{1}s_{1}+m_{2}l_{2}s_{1+2}\\
m_{2}l_{2}s_{1+2}
\end{array}\right]$
\par\end{center}

Finally, we have a set of first-order differential equations, which
we can throw into some numerical integration solver. Here we will
use the built-in solver from MATLAB. The following two figures are
frames from the simulation of a passive acrobot swing from right to
left.
\begin{center}
\includegraphics[width=100pt,height=100pt]{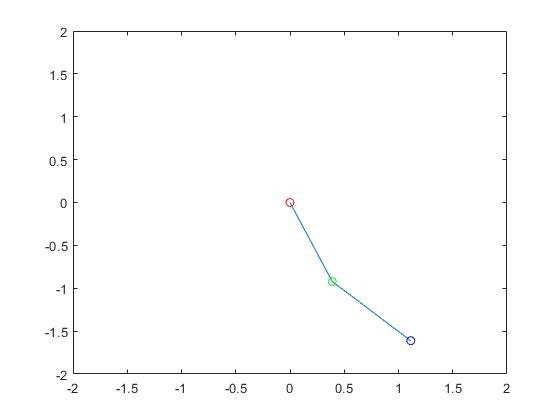}\includegraphics[width=100pt,height=100pt]{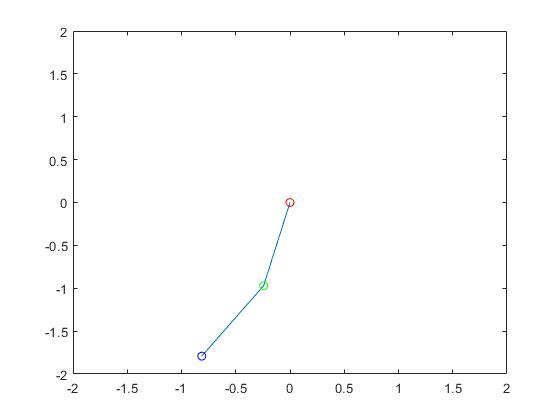}
\par\end{center}

\subsection{Trajectory Optimization}

Alright we have a passive double pendulum swinging back and forth,
but that is not actually very exciting. The exciting part of this
note is to demonstrate how to swing up the acrobot from the bottom
using a control technique called trajectory optimization. The software
library we are going to use is an open-sourced trajectory optimization
library called OptimTraj.

Now, we have to reformulate the equations of motion into the cannonical
form, so our plan is:
\begin{center}
$\dot{q}=Mass(q)^{-1}f(q,u)$,
\par\end{center}

which is a non-linear dynamical system. Then, we can construct an
optimization problem using this model, which we can throw into our
trajectory optimization library. The following fugure plots the trajectories
of the two joints(red and green joints).
\begin{center}
\includegraphics[height=150pt]{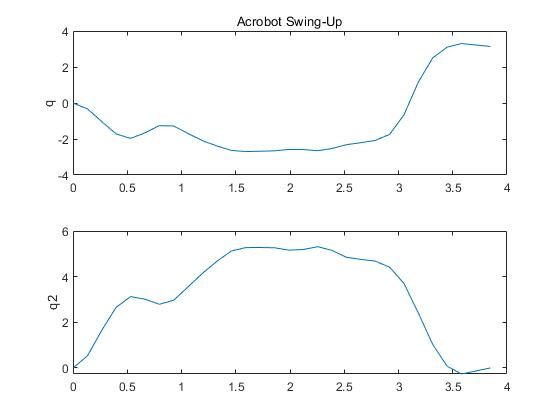}
\par\end{center}

The values of these two actuated joints start from $[0,0]$ and end
at $[\pi,0]$. Then, let's animate the swing up sequence of the 2-link
acrobot. Here are three frames from the animation.
\begin{center}
\includegraphics[width=100pt,height=100pt]{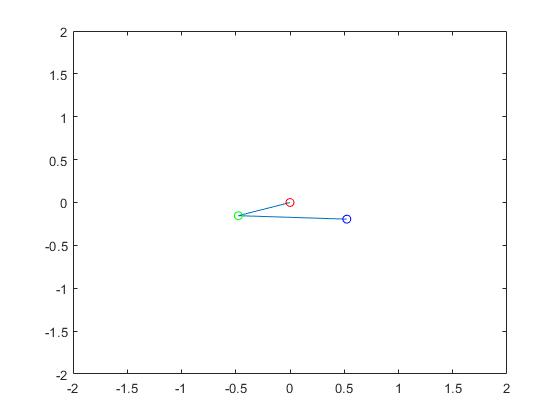}\includegraphics[width=100pt,height=100pt]{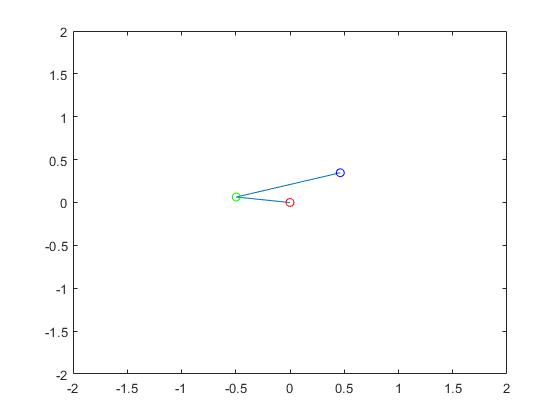}\includegraphics[width=100pt,height=100pt]{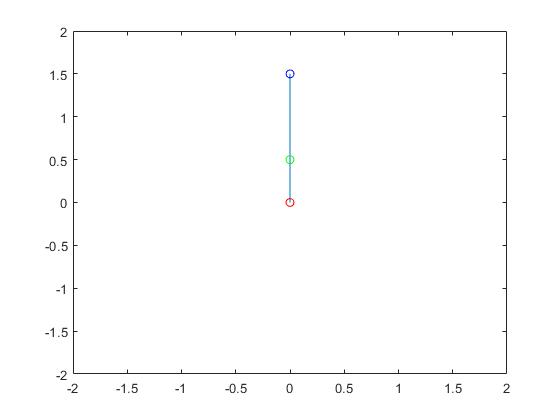}
\par\end{center}

\section{A 3-link Acrobot}

\subsection{Equations of Motion}

Let's derive the equations of motion for a 3-link acrobot in 2-dimensional
space. Here we are again using the method of Lagrange. Let's get start
with the Lagrangian, which comes from the kinetic and potential energy
of the system.
\begin{center}
$T=\frac{1}{2}(m_{1}+m_{2}+m_{3})l_{1}^{2}\dot{q}_{1}^{2}+\frac{1}{2}(m_{2}+m_{3})l_{2}^{2}(\dot{q_{1}}+\dot{q_{2}})^{2}+\frac{1}{2}m_{3}l_{3}^{2}(\dot{q_{1}}+\dot{q_{2}}+\dot{q_{3}})^{2}+(m_{2}+m_{3})l_{1}l_{2}\dot{q_{1}}(\dot{q_{1}}+\dot{q_{2}})c_{2}+m_{3}l_{1}l_{3}\dot{q_{1}}(\dot{q_{1}}+\dot{q_{2}}+\dot{q_{3}})c_{2+3}+m_{3}l_{2}l_{3}(\dot{q_{1}}+\dot{q_{2}})(\dot{q_{1}}+\dot{q_{2}}+\dot{q_{3}})c_{3}$
\par\end{center}

\begin{center}
$U=-(m_{1}+m_{2}+m_{3})l_{1}gc_{1}-(m_{2}+m_{3})l_{2}gc_{1+2}-m_{3}l_{3}gc_{1+2+3}$
\par\end{center}

\begin{center}
$L=T-U$
\par\end{center}

Next, let's recall our favorite Lagrange Equations.
\begin{center}
$\frac{d}{dt}\frac{\partial L}{\partial\dot{q_{i}}}-\frac{\partial L}{\partial q_{i}}=\tau_{i}$
\par\end{center}

Plugging the Lagrangian into these equations yields the equations
of motion.
\begin{center}
$\tau_{1}=(m_{1}+m_{2}+m_{3})l_{1}^{2}\ddot{q_{1}}+(m_{2}+m_{3})l_{2}^{2}(\ddot{q_{1}}+\ddot{q_{2}})+m_{3}l_{3}^{2}(\ddot{q_{1}}+\ddot{q_{2}}+\ddot{q_{3}})+(m_{2}+m_{3})l_{1}l_{2}(2\ddot{q_{1}}+\ddot{q_{2}})c_{2}-(m_{2}+m_{3})l_{1}l_{2}(2\dot{q_{1}}+\dot{q_{2}})\dot{q_{2}}s_{2}+m_{3}l_{1}l_{3}(2\ddot{q_{1}}+\ddot{q_{2}}+\ddot{q_{3}})c_{2+3}-m_{3}l_{1}l_{3}(2\dot{q_{1}}+\dot{q_{2}}+\dot{q_{3}})(\dot{q_{2}}+\dot{q_{3}})s_{2+3}+m_{3}l_{2}l_{3}(2\ddot{q_{1}}+2\ddot{q_{2}}+\ddot{q_{3}})c_{3}-m_{3}l_{2}l_{3}(2\dot{q_{1}}+\dot{q_{2}}+\dot{q_{3}})\dot{q_{3}}s_{3}+(m_{1}+m_{2}+m_{3})l_{1}gs_{1}+(m_{2}+m_{3})l_{2}gs_{1+2}+m_{3}l_{3}gs_{1+2+3}$
\par\end{center}

\begin{center}
$\tau_{2}=(m_{2}+m_{3})l_{2}^{2}(\ddot{q_{1}}+\ddot{q_{2}})+m_{3}l_{3}^{2}(\ddot{q_{1}}+\ddot{q_{2}}+\ddot{q_{3}})+(m_{2}+m_{3})l_{1}l_{2}\ddot{q_{1}}c_{2}-(m_{2}+m_{3})l_{1}l_{2}\dot{q_{1}}\dot{q_{2}}s_{2}+m_{3}l_{1}l_{3}\ddot{q_{1}}c_{2+3}-m_{3}l_{1}l_{3}\dot{q_{1}}(\dot{q_{2}}+\dot{q_{3}})s_{2+3}+m_{3}l_{2}l_{3}(2\ddot{q_{1}}+2\ddot{q_{2}}+\ddot{q_{3}})c_{3}-m_{3}l_{2}l_{3}(2\dot{q_{1}}+2\dot{q_{2}}+\dot{q_{3}})\dot{q_{3}}s_{3}+(m_{2}+m_{3})l_{1}l_{2}\dot{q_{1}}(\dot{q_{1}}+\dot{q_{2}})s_{2}+m_{3}l_{1}l_{3}(\dot{q_{1}}+\dot{q_{2}})(\dot{q_{1}}+\dot{q_{2}}+\dot{q_{3}})s_{2}+(m_{2}+m_{3})l_{2}gs_{1+2}+m_{3}l_{3}gs_{1+2+3}$
\par\end{center}

\begin{center}
$\tau_{3}=m_{3}l_{3}^{2}(\ddot{q_{1}}+\ddot{q_{2}}+\ddot{q_{3}})+m_{3}l_{1}l_{3}\ddot{q_{1}}c_{2+3}-m_{3}l_{1}l_{3}\dot{q_{1}}(\dot{q_{2}}+\dot{q_{3}})s_{2+3}+m_{3}l_{2}l_{3}(\ddot{q_{1}}+\ddot{q_{2}})c_{3}-m_{3}l_{2}l_{3}(\dot{q_{1}}+\dot{q_{2}})\dot{q_{3}}s_{3}+m_{3}l_{1}l_{3}\dot{q_{1}}(\dot{q_{1}}+\dot{q_{2}}+\dot{q_{3}})s_{2+3}+m_{3}l_{2}l_{3}(\dot{q_{1}}+\dot{q_{2}})(\dot{q_{1}}+\dot{q_{2}}+\dot{q_{3}})s_{3}+m_{3}l_{3}gs_{1+2+3}$
\par\end{center}

\subsection{Dynamics Model of 2-link Acrobot}

These equations may look horrific for normal people(include me). However,
we can put them into a nicer looking matrix form, which is what we
have to tell the computer. The Manipulator Equations of our 3-link
acrobot takes the form:

$M(q)\ddot{q}+C(q,\dot{q})\dot{q}=\tau_{g}(q)+Bu$

We have to again reformulate the problem into a first-order differential
equation by constructing the mass matrix. This time, the mass matrix
is a $6\times6$ matrix, since our state variable has became 6-dimensional.
\begin{center}
$Mass(q)=\left[\begin{array}{cc}
I_{3\times3} & 0_{3\times3}\\
C(q)_{3\times3} & M(q)_{3\times3}
\end{array}\right]$
\par\end{center}

As expected, the mass matrix is going to be a little bit complicated,
so we decide to present the matrix element-by-element. $M_{ij}$ denotes
elements from matrix $M(q)$, while $C_{ij}$ denotes elements from
matrix $C(q,\dot{q})$.
\begin{center}
$M_{11}=(m_{1}+m_{2}+m_{3})l_{1}^{2}+(m_{2}+m_{3})l_{2}^{2}+m_{3}l_{3}^{2}+2(m_{2}+m_{3})l_{1}l_{2}c_{2}+2m_{3}l_{1}l_{3}c_{2+3}+2m_{3}l_{2}l_{3}c_{3}$
\par\end{center}

\begin{center}
$M_{12}=(m_{2}+m_{3})l_{2}^{2}+m_{3}l_{3}^{2}+(m_{2}+m_{3})l_{1}l_{2}c_{2}+m_{2}l_{1}l_{3}c_{2+3}+2m_{3}l_{2}l_{3}c_{3}$
\par\end{center}

\begin{center}
$M_{13}=m_{3}l_{3}^{2}+m_{3}l_{1}l_{3}c_{2+3}+m_{3}l_{2}l_{3}c_{3}$
\par\end{center}

\begin{center}
$M_{21}=(m_{2}+m_{3})l_{2}^{2}+m_{3}l_{3}^{2}+(m_{2}+m_{3})l_{1}l_{2}c_{2}+m_{3}l_{1}l_{3}c_{2+3}+2m_{3}l_{2}l_{3}c_{3}$
\par\end{center}

\begin{center}
$M_{22}=(m_{2}+m_{3})l_{2}^{2}+m_{3}l_{3}^{2}+2m_{3}l_{2}l_{3}c_{3}$
\par\end{center}

\begin{center}
$M_{23}=m_{3}l_{3}^{2}+m_{3}l_{2}l_{3}c_{3}$
\par\end{center}

\begin{center}
$M_{31}=m_{3}l_{3}^{2}+m_{3}l_{1}l_{3}c_{2+3}+m_{3}l_{2}l_{3}c_{3}$
\par\end{center}

\begin{center}
$M_{32}=m_{3}l_{3}^{2}+m_{3}l_{2}l_{3}c_{3}$
\par\end{center}

\begin{center}
$M_{33}=m_{3}l_{3}^{2}$
\par\end{center}

\begin{center}
$C_{11}=0$
\par\end{center}

\begin{center}
$C_{12}=-(m_{2}+m_{3})l_{1}l_{2}(2\dot{q_{1}}+\dot{q_{2}})s_{2}-m_{3}l_{1}l_{3}(2\dot{q_{1}}+\dot{q_{2}}+\dot{q_{3}})s_{2+3}$
\par\end{center}

\begin{center}
$C_{13}=-m_{3}l_{1}l_{3}(2\dot{q_{1}}+\dot{q_{2}}+\dot{q_{3}})s_{2+3}-m_{3}l_{2}l_{3}(2\dot{q_{1}}+\dot{q_{2}}+\dot{q_{3}})s_{3}$
\par\end{center}

\begin{center}
$C_{21}=-(m_{2}+m_{3})l_{1}l_{2}\dot{q_{2}}s_{2}-m_{3}l_{1}l_{3}(\dot{q_{2}}+\dot{q_{3}})s_{2+3}+(m_{2}+m_{3})l_{1}l_{2}(\dot{q_{1}}+\dot{q_{2}})s_{2}+m_{3}l_{1}l_{3}(\dot{q_{1}}+\dot{q_{2}}+\dot{q_{3}})s_{3}$
\par\end{center}

\begin{center}
$C_{22}=m_{3}l_{1}l_{3}(\dot{q_{1}}+\dot{q_{2}}+\dot{q_{3}})s_{3}$
\par\end{center}

\begin{center}
$C_{23}=-m_{3}l_{2}l_{3}(2\dot{q_{1}}+2\dot{q_{2}}+\dot{q_{3}})s_{3}$
\par\end{center}

\begin{center}
$C_{31}=-m_{3}l_{1}l_{3}(\dot{q_{2}}+\dot{q_{3}})s_{2+3}+m_{3}l_{1}l_{3}(\dot{q_{1}}+\dot{q_{2}}+\dot{q_{3}})s_{2+3}+m_{3}l_{2}l_{3}(\dot{q_{1}}+\dot{q_{2}}+\dot{q_{3}})s_{3}$
\par\end{center}

\begin{center}
$C_{32}=m_{3}l_{2}l_{3}(\dot{q_{1}}+\dot{q_{2}}+\dot{q_{3}})s_{3}$
\par\end{center}

\begin{center}
$C_{33}=-m_{3}l_{2}l_{3}(\dot{q_{1}}+\dot{q_{2}})s_{3}$
\par\end{center}

Very cool! Now we are ready to put everything into the computer. In
MATLAB, again we construct the following first-order differential
equations:
\begin{center}
$\left[\begin{array}{cc}
I_{3\times3} & 0_{3\times3}\\
C(q)_{3\times3} & M(q)_{3\times3}
\end{array}\right]\left[\begin{array}{c}
q_{1}\\
q_{2}\\
q_{3}\\
\dot{q_{1}}\\
\dot{q_{2}}\\
\dot{q_{3}}
\end{array}\right]=\left[\begin{array}{c}
\dot{q_{1}}\\
\dot{q_{2}}\\
\dot{q_{3}}\\
\tau_{g1}+u_{1}\\
\tau_{g1}+u_{2}\\
\tau_{g3}+u_{3}
\end{array}\right]$,
\par\end{center}

where $\tau_{gi}$ are the gravity terms, while $u_{i}$ are the control
input. Here we have three input, which means the system is fully actuated.
Since the acrobot is underactuated, we will take away $u_{1}$ when
we solve the acrobot problem. The gravity terms are:
\begin{center}
$\tau_{g1}=-g\left[(m_{1}+m_{2}+m_{3})l_{1}s_{1}+(m_{2}+m_{3})l_{2}s_{1+2}+m_{3}l_{3}s_{1+2+3}\right]$
\par\end{center}

\begin{center}
$\tau_{g2}=-g\left[(m_{2}+m_{3})l_{2}s_{1+2}+m_{3}l_{3}s_{1+2+3}\right]$
\par\end{center}

\begin{center}
$\tau_{g3}=-gm_{3}l_{3}s_{1+2+3}$
\par\end{center}

Exciting news! All the mathematic is done and now let's talk about
how to do simulation on a computer.

The first thing we want to work out on the computer is to check whether
we have the correct dynamics model. To check whether our model is
correct, an intuitive way is simulate the passive system. We can give
the system an initial state, such as $[\begin{array}{cccccc}
\frac{\pi}{3} & 0 & 0 & 0 & 0 & 0\end{array}]$, and watch it evolves over time. If there are any bugs in our equations,
chances are the system will looked unreal. I have to clarify that
this method doesn't gauruntee the model is correct, but I choose to
believe my eyeballs and commonsense(because I'm an engineer, typically
lazy person). Since our model is barely a set of differential equations,
we can use numerical integration tools to solve it approximatly. Our
plan here is using the built-in solver from MATLAB. The dynamics model
of acrobot fits right in one of the standard form support by MATLAB,
which called ``ODE with time and state dependent mass matrix''.
In order to tell MATLAB what we intended to do, we have to specify
options for the ode45 solver.

$options=odeset('Mass',@mass)$,

$[t,y]=ode45(@f,tspan,q0,options)$,

where the name/value pair $('Mass',@mass)$ tells ode45 that we have
a mass matrix in our equations and we have a function called 'mass'
calculating it. In the second line of code, 'tspan' stands for the
time span we wish to integrate; 'q0' is the initial state of the system;
'@f' link ode45 to a function called 'f', which calculates the time
derivative of the state variables.

The ode45 solver will give us the numerical solution of our system
over time, given our manually picked initial state. Then we can plot
the solution and convince ourselves we have the correct model, if
we are lucky to be typo-free. In the following images, the acrobot
swing from right to left passively.
\begin{center}
\includegraphics[width=100pt,height=100pt]{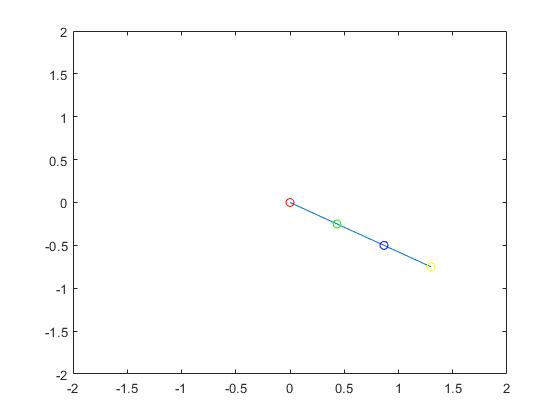}\includegraphics[width=100pt,height=100pt]{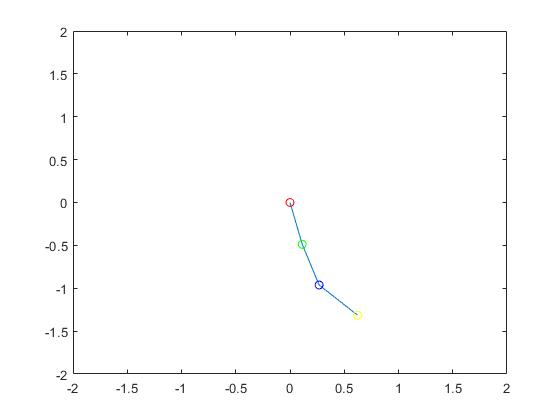}\includegraphics[width=100pt,height=100pt]{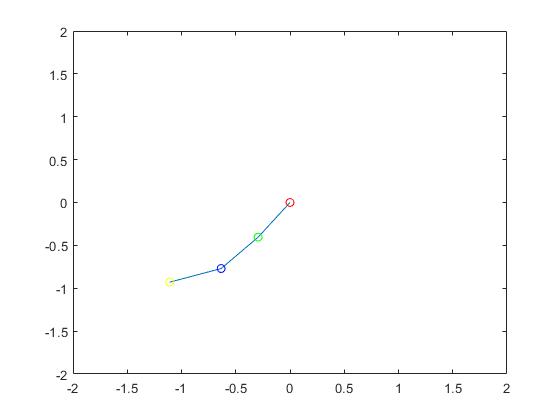}
\par\end{center}

\subsection{Trajectory Optimization}

Finally we are here, after all the ugly mathematics and boring simulation.
We have all we need to wirte a piece of code to solve the 3-link acrobot
swing-up problem. We have a dynamics model of the system, which we've
convined ourselves is correct, and a library to perform trajectory
optimization. Now, we will construct an trajectory optimization using
the dynamics model and throw it to OptimTraj.

To use the OptimTraj library, we need to first construct a function
which computes the dynamics. Here we have to put the inverse of the
mass matrix on the right hand side, like

$\dot{q}=Mass(q)^{-1}f(q,u)$,

where the inverse matrix is provable to exist. The second thing is
to supply a cost function, which we will take the dot product of the
control effort. This will leads to the minimization of control effort.
Then we also need to specify some constraint for the optimization
problem, including initial state, final state, upper and lower bounds
of state variables and initial guesses. After specifying all these
stuff, we finally can leave the hard work to computer. The following
two figures plot the trajectories of two actuated joints(green and
blue joints).
\begin{center}
\includegraphics[height=150pt]{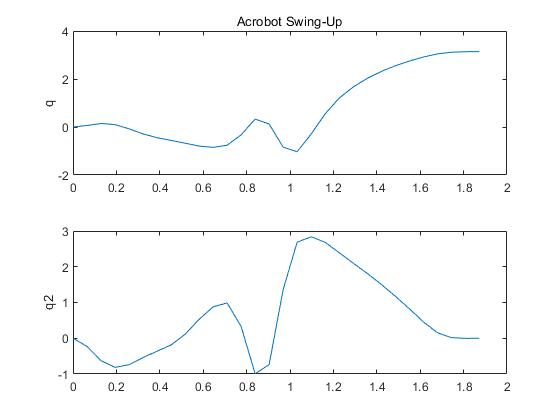}
\par\end{center}

The values of these two actuated joints start from $[0,0]$ and end
at $[\pi,0]$. Then, let's animate the swing up sequence of the 3-link
acrobot. Here are three frames from the animation.
\begin{center}
\includegraphics[width=100pt,height=100pt]{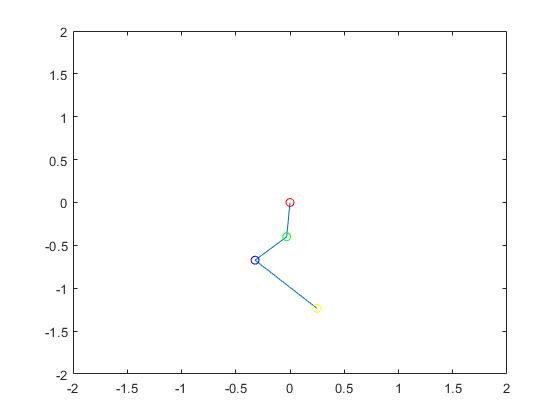}\includegraphics[width=100pt,height=100pt]{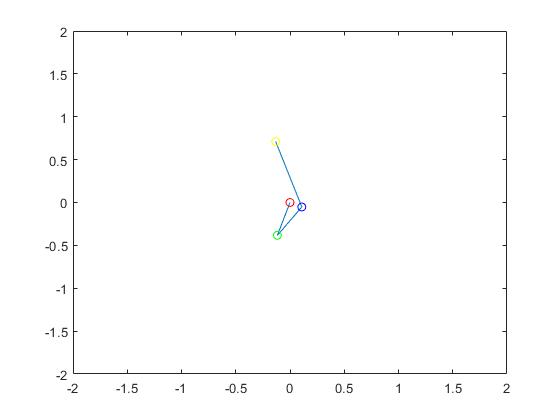}\includegraphics[width=100pt,height=100pt]{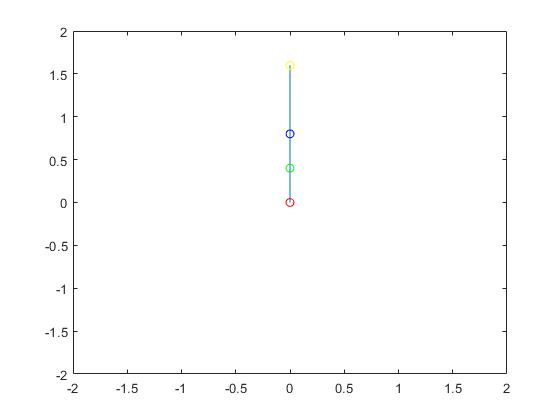}
\par\end{center}

\section{Remarks}

This note present an experimental study of acrobot, which is a easygoing
member in the family of underactuated robots. To make the result mentioned
here concrete, stability and controlability analysis are expected
in the future.

\end{document}